\documentclass[fleqn,10pt]{wlscirep}

\usepackage{graphicx}
\usepackage{amsfonts}
\usepackage{amsmath}
\usepackage{mathtools}
\usepackage{float}
\usepackage{epsfig}
\usepackage{color}
\usepackage{gensymb}
\usepackage{multirow}
\usepackage[mathscr]{euscript}

\newcommand{\onlinecite}[1]{\hspace{-1 ex} \nocite{#1} \citenum{#1}} 
\newcommand{\e}[1]{\ensuremath{\times 10^{#1}}}

\title{The Perfect Glass Paradigm: Disordered Hyperuniform Glasses Down to Absolute Zero}

\author[1]{G. Zhang}
\author[1]{F. H. Stillinger}
\author[2,*]{S. Torquato}
\affil[1]{Department of Chemistry, Princeton University, Princeton, 08540, USA}
\affil[2]{Department of Chemistry, Department of Physics,
Princeton Institute for the Science and Technology of
Materials, and Program in Applied and Computational Mathematics, Princeton University, Princeton, 08540, USA}

\affil[*]{torquato@electron.princeton.edu}

\begin{abstract}

Rapid cooling of liquids below a certain temperature range can result in a transition to glassy states. The traditional understanding of glasses includes their thermodynamic metastability with respect to crystals. 
However, here we present specific examples of interactions that eliminate the possibilities of crystalline and quasicrystalline phases, while creating mechanically stable amorphous glasses down to absolute zero temperature. We show that this can be accomplished by introducing a new ideal state of matter called a ``perfect glass.'' A perfect glass represents a soft-interaction analog of the maximally random jammed (MRJ) packings of hard particles.  These latter states can be regarded as the epitome of a glass since they are out of equilibrium, maximally disordered, hyperuniform, mechanically rigid with infinite bulk and shear moduli, and can never crystallize due to configuration-space trapping. Our model perfect glass utilizes two-, three-, and four-body soft interactions while simultaneously retaining the salient attributes of the MRJ state. 
These models constitute a theoretical proof of concept for perfect glasses and broaden our fundamental understanding of glass physics. A novel feature of equilibrium systems of identical particles interacting with the perfect-glass potential at positive temperature is that they have a non-relativistic speed of sound that is infinite.

\end{abstract}
\begin{document}

\flushbottom
\maketitle
%
%
\thispagestyle{empty}

\section*{Introduction}

\begin{figure}
\begin{center}
\includegraphics[width=0.45\textwidth]{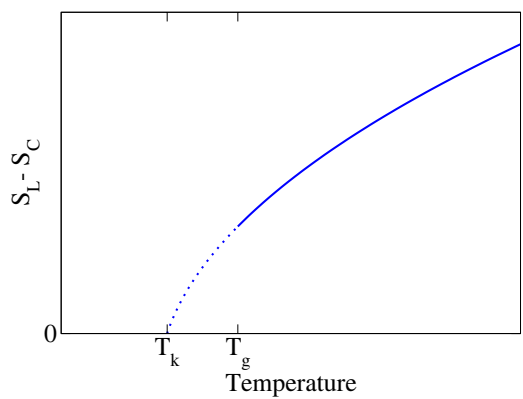}

\end{center}
\caption{Schematic illustration of the so-called Kauzmann paradox. As a liquid is supercooled, the entropy difference between it and crystalline state, $S_L-S_C$, decreases. If the glass transition can be postponed below the Kauzmann temperature, $T_k$, then the entropy of the liquid would be lower than that of the crystal upon extrapolation. The perfect-glass paradigm introduced in this paper completely circumvents the Kauzmann paradox.}
\label{kauzmann}
\end{figure}

Structural glasses are materials made by supercooling liquids below the ``glass transition temperature,'' sufficiently rapidly to avoid crystallization \cite{angell1988perspective}.
According to Ref.~\onlinecite{chaikin2000principles}, a qualitative description of a structural glass is ``a phase of matter with no long-range order but with a nonzero shear rigidity.''
It is well known that the glass transition temperature can be reduced by lowering the cooling rate. However, some have postulated that if the glass transition temperature could be postponed down to absolute zero during the supercooling process, then at some low but positive temperature, called the ``Kauzmann temperature,'' the entropy of the supercooled liquid would be equal to and then apparently decline below that of the crystal, resulting in the so-called ``Kauzmann paradox'' \cite{kauzmann1948nature, xu2016entropy}, which is schematically depicted in Fig.~\ref{kauzmann}.
One resolution of this well-known paradox is to assume that supercooled liquids at the Kauzmann temperature must undergo a thermodynamic phase transition to ``ideal glasses''. Such glasses identified as ideal would have vanishing extensive configurational entropy \cite{debenedetti2001supercooled, stillinger2013glass}. In this paper, however, we present a completely different model system that we call a  ``perfect glass.'' As we will see, one characteristic of the perfect-glass paradigm introduced in the present paper is the complete circumvention of the Kauzmann paradox.

Various studies have justifiably placed importance on the local environment of each atom in structural glasses \cite{ma2015tuning, sheng2006atomic, starr2002we}.
The variations in local motifs and the resulting varying degrees of short-range order have been used to explain the physical properties of glasses. For example, it is believed that atoms in ordered local environments are responsible for thermodynamic properties and kinetic stability of glasses, while atoms in disordered local environments make important contributions to mechanical properties \cite{ma2015tuning}.
By contrast, global structural characteristics, such as hyperuniformity, as described below, address key aspects of glass formation that have been unexplored by these local descriptive techniques.

Although the compositions and interactions of experimentally investigated glasses are generally complicated on the atomic scale, many theoretical and computational models with simpler compositions and interactions have been shown to produce glasses under rapid cooling \cite{kristensen1976computer, sciortino2004equilibrium}. Moreover, short-range, pairwise additive interactions have been specifically designed to locally frustrate crystallization to create good glass formers \cite{dzugutov1992glass, kob1994scaling, perera1999relaxation, shintani2006frustration, jonsson1988icosahedral, wales2003energy}. This is often achieved by having two
components
whose simultaneous existence disrupts crystal nucleation in two dimensions (2D) \cite{kob1994scaling, perera1999relaxation} or three dimensions (3D) \cite{jonsson1988icosahedral}, or by encouraging pentagonal or icosahedral local geometry  \cite{roth2000solid, steinhardt1983bond}  that frustrates crystallization  \cite{dzugutov1992glass, shintani2006frustration}.
Despite these design goals to strongly inhibit crystal nucleation, the true ground states of these potentials nevertheless turn out to be crystalline \cite{roth2000solid, fernandez2003crystal, perera1999stability, shintani2006frustration}. Therefore, liquids with these interactions can still crystallize if cooled slowly enough. 
Interestingly, there are models that produce amorphous ground states \cite{uche2004constraints, martinez2011design, smallenburg2013liquids},
but they cannot resist shear and 
hence do not behave like glasses, which are mechanically stable.
Moreover, in all of these cases,
 crystalline structures are
still part of the ground-state manifold, 
even if the probability of observing them is extremely small.
Therefore, placing such systems in contact with suitable periodic substrates would have the effect of inducing crystallization with an appreciably higher probability.
As we will see, the perfect glass paradigm does not even allow this to occur because ordered states (for all temperatures) are completely banished.

Maximally-random jammed (MRJ) packings of hard (nonoverlapping) particles in 2D and 3D are idealized amorphous states of matter that can be regarded to be 
prototypical glasses \cite{torquato2000random, zachary2011hyperuniform, jiao2011maximally, torquato2010jammed}.
A packing is called ``strictly jammed'' if no subset of particles may
be displaced while allowing uniform volume-preserving deformations
of the system boundary \cite{torquato2003breakdown}, implying resistance to both
compressive and shear deformations. Among all strictly jammed packings, MRJ states are defined to be the most disordered ones according to  suitable order metrics (i.e.,  measures of the degree of geometric order).
MRJ packings not only exhibit many characteristics that are typical of glasses, but also are extremal in several respects according to the description
given in Ref.~\onlinecite{chaikin2000principles}: They are nonequilibrium, nonergodic many-body systems that are maximally disordered subject to the nonoverlap constraint, non-crystallizable, and mechanically infinitely rigid (both elastic moduli are unbounded) \cite{torquato2003breakdown}.
Indeed, they are perfectly nonergodic, since they are forever trapped in configuration space.

However, there are still two major differences between MRJ packings and molecular glasses.
First, MRJ packings are hyperuniform \cite{torquato2003local, donev2005unexpected, zachary2011hyperuniform, kurita2011incompressibility, jiao2011maximally}, while typical molecular glasses are not \cite{marcotte2013nonequilibrium}.
A hyperuniform many-particle system is one in which the structure factor approaches zero in the infinite-wavelength limit \cite{torquato2003local}. In such systems, density fluctuations are anomalously suppressed at very large length scales \cite{torquato2003local}, which imposes strong global structural constraints. All structurally perfect crystals are hyperuniform, but typical disordered many-particle systems, including liquids and molecular glasses, are not. Materials that are simultaneously disordered and hyperuniform can be regarded to be exotic states of matter that lie between a crystal and a liquid; they behave more like crystals in the manner in which they
suppress large-scale density fluctuations, and yet they also
resemble typical statistically isotropic liquids with no Bragg peaks, and hence have no long-range order.
Therefore, disordered hyperuniform states of matter have been the subject of many recent investigations \cite{torquato2003local, donev2005unexpected, zachary2011hyperuniform, jiao2011maximally, kurita2011incompressibility, dreyfus2015diagnosing, lesanovsky2014out, hexner2015hyperuniformity, jack2015hyperuniformity, de2015toward, degl2015thz, xie2013hyperuniformity, muller2014silicon, florescu2009designer}.

Second, given sufficiently long observation times at positive temperature, a typical molecular glass will eventually crystallize because the free energy barrier between it and its corresponding stable crystal structure is finite. By contrast, the hard-sphere MRJ state is a singular point that is trapped in a jamming basin in configuration space and hence can never crystallize at constant volume \cite{donev2004linear}.

\begin{figure}
\begin{center}
\includegraphics[width=0.7\textwidth]{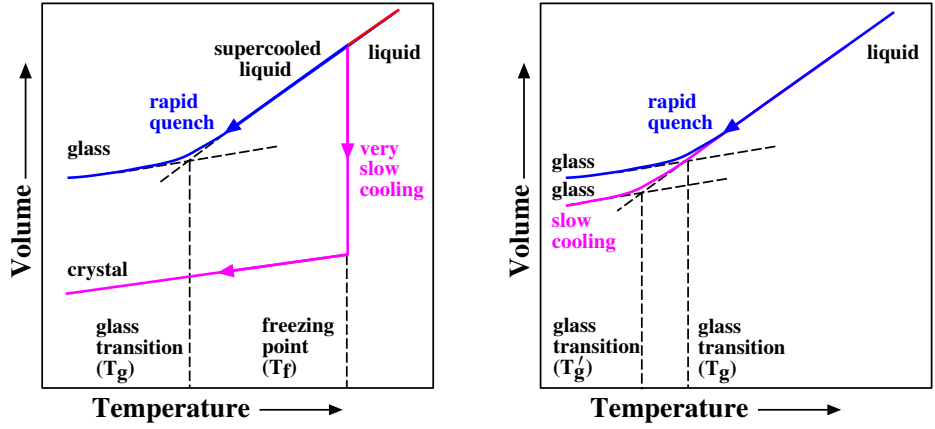}
\end{center}
\caption{Schematic constant-pressure phase diagrams. Left panel: Typical many-particle systems become glasses upon rapid cooling or can crystallize upon slow cooling. Right panel: Our model family creates perfect glasses that by construction cannot crystallize upon quenching to absolute zero temperature.}
\label{phasediagram}
\end{figure}

Geometrically motivated by MRJ extremal glasses, we are interested in constructing molecular-glass analogs (that are not limited to pairwise additive interactions), which we call ``perfect glasses''\footnote{Our definition of perfect glass is distinctly different from the ``ideal glass former'' with reversed $T_g$ and $T_m$ \cite{kapko2013ideal}.}. These analogs should exhibit the following attributes: (1) be disordered and hyperuniform (a global criterion as opposed to the local coordination geometry focus of previous studies);
(2) possess no crystalline or quasicrystalline energy minima so that they remain disordered even in the infinite-observation-time limit at positive  temperature, implying that they can
never crystallize (in contrast to conventional glass formers), as qualitatively shown
in Fig.~\ref{phasediagram}; and
(3) must possess both positive bulk and shear moduli.

In this paper, we explicitly show that such perfect glasses mathematically exist. Specifically, we demonstrate that a single-component system with a combination of long-ranged two-, three-, and four-body interactions can produce perfect glasses. Our perfect glass interactions are designed in Fourier space which allows us independently to tune the structure factor over the entire range from infinite to intermediate wavelength, including values that will automatically include all possible Bragg peaks, while maintaining hyperuniformity. These global constraints therefore permit the suppression of all possible Bragg peaks, which by definition eliminates any crystal and quasicrystal formation. This attribute of the perfect glass stands in contrast with the ideal glass concept \cite{debenedetti2001supercooled, stillinger2013glass} as well as with previous studies that are intended to frustrate crystallization via local coordination preferences \cite{dzugutov1992glass, kob1994scaling, perera1999relaxation, shintani2006frustration}. In the case of the former, this means that there is no entropy catastrophe because there is no ordered structure with which to carry out an entropy comparison. Our global approach of preventing crystallization applies to any dimension,  in contrast to previously designed interactions that are specifically tailored for a particular space dimension \cite{dzugutov1992glass, kob1994scaling, perera1999relaxation, shintani2006frustration}.

The major features of perfect glasses are not limited to our three criteria (hyperuniform, possess no crystalline or quasicrystalline energy minima, and possess both positive bulk and shear moduli). We will also see that  perfect glasses completely circumvent Kauzmann's paradox. Besides the theoretical existence of perfect glasses, another important finding of our work is that liquid-state configurations of our perfect-glass interaction are hyperuniform and hence possess a zero internal compressibility, implying they have a non-relativistic speed of sound that is infinite.
Finally, our results also suggest that up to four-body interactions are necessary to completely avoid crystallization, and thus explains the failure to create such an ideal state of matter heretofore.

\section*{Perfect Glass Potentials}
\label{sec_potential}

We apply the collective-coordinate
optimization scheme \cite{uche2004constraints, zhang2015ground} to construct
interactions that can produce perfect
glasses. This procedure involves
finding potentials 
that are given in terms of a targeted
form of the structure factor. For a single-component system with $N$ particles in a simulation box of volume $V$ with periodic boundary conditions in $d$-dimensional Euclidean space $\mathbb R^d$, the single-configuration structure factor is defined as $\mathscr S(\mathbf k)=|\sum_{j=1}^N \exp(-i \mathbf k \cdot \mathbf r_j)|^2/N$, where $\mathbf k$ is a $d$-dimensional wavevector and $\mathbf r_j$ is the position of particle $j$ \cite{chandler1987introduction, chaikin2000principles}.
Many previous investigations have focused on targeting ``stealthy"
structure factors, i.e., those in which
the structure factor is exactly zero
within some sphere of radius $K$ around the origin
in Fourier space \cite{uche2004constraints, suto2005crystalline, torquato2015ensemble, zhang2015ground, zhang2015ground2}. 
When $K$
is large the ground states are crystalline \cite{uche2004constraints, suto2005crystalline, torquato2015ensemble, zhang2015ground}.
When $K$
is sufficiently small, however, it has been shown
that the ground states of the associated  interactions are disordered and highly degenerate
\cite{uche2004constraints, torquato2015ensemble, zhang2015ground}.
However, although these states are hyperuniform, they are not perfect glasses because they cannot resist shear
and crystal structures are part of the 
ground-state manifold, even if they
are sets of zero measure in the infinite-volume limit.

However, the collective-coordinate optimization scheme has also been used to prescribe the potential
energy $\Phi$ defined by the following more general targeted structure factor \cite{uche2006collective,batten2008classical, zachary2011anomalous}:
\begin{equation}
\Phi(\mathbf r^N) = \sum_{0<|\mathbf k|<K} {\tilde v}(\mathbf k) [\mathscr S(\mathbf k)-\mathscr S_0(\mathbf k)]^2,
\label{Potential_Fourier}
\end{equation}
where $\mathbf r^N=\mathbf r_1,\mathbf r_2,...,\mathbf r_N$ represents the configurational coordinates, the summation is over all reciprocal lattice vector $\mathbf k$'s of the simulation box \cite{chaikin2000principles} such that $0<|\mathbf k|<K$, $\mathscr S_0(\mathbf k)$ is a ``target'' structure factor, ${\tilde v}(\mathbf k)>0$ is a weight function, and $K$ is some cut-off
wavenumber that determines the number of constrained wave vectors. At low temperature, this interaction potential attempts to ``constrain'' the structure factor $\mathscr S(\mathbf k)$ to the target $\mathscr S_0(\mathbf k)$ for all $|\mathbf k|<K$, since violating a constraint for any $\mathbf k$ will increase the potential energy. The number of independent\footnote{The definition of $\mathscr S(\mathbf k)$ implies that $\mathscr S(\mathbf k) = \mathscr S(-\mathbf k)$. Therefore, not all constraints are independent.} constraints divided by the total number of degrees of freedom, $d(N-1)$, is a parameter that measures how constrained the system is and is denoted by $\chi$. Previous research has focused on $\chi$ values less than 1. In such under-constrained cases, a minimum of the potential energy satisfies all constraints while still having leftover unconstrained degrees of freedom.
Although this interaction is defined in Fourier space (i.e., in terms of the structure factor), it can be decomposed into a sum of two-body, three-body, and four-body terms in direct space \cite{uche2006collective}. In Appendix, we present explicit formulas for each term. In the Supplementary Information (SI), visualizations of these contributions to the potential energy are provided.

For several reasons, such a model is an excellent starting point for designing perfect glass interactions. First, this model enables one to fulfill the requirement that perfect glasses be hyperuniform because this model constrains $\mathscr S(\mathbf k)$ to a targeted hyperuniform functional form $\mathscr S_0(\mathbf k)$ around the origin. In this paper, we select the following form for $\mathscr S_0(\mathbf k)$:
\begin{equation}
\mathscr S_0(\mathbf k)=|\mathbf k|^\alpha \mbox{ ~~ for }0 \le k \le K,
\label{target}
\end{equation}
where $\alpha>0$ is an exponent that we are free to prescribe. To ensure that $\mathscr S(\mathbf k)$ has the targeted hyperuniform power-law form of $\mathscr S_0(\mathbf k)$, we choose a weight function that diverges at the origin:
\begin{equation}
{\tilde v}(\mathbf k) = \left(\frac{1}{|\mathbf k|}-1\right)^\gamma \mbox{ ~~ for }0 \le k \le K,
\label{weight}
\end{equation}
where $\gamma \ge 2$ is another exponent to choose. The choices of $\mathscr S_0(\mathbf k)$ and ${\tilde v}(\mathbf k)$ are not unique: Other target forms of $\mathscr S_0(\mathbf k)$ and other forms of ${\tilde v}(\mathbf k)$ that diverge to $+\infty$ in the zero wavenumber limit could also result in hyperuniformity. We choose the forms in Eqs.~(\ref{target})-(\ref{weight}) for simplicity.
Because ${\tilde v}(\mathbf k)$ goes to zero smoothly as $|\mathbf k|$ goes to 1, it is natural to let $K=1$. Our choice of $K$ and ${\tilde v}(\mathbf k)$ sets the model's length and energy scales.

Second, this model can completely eliminate crystalline and quasicrystalline energy minima. The structure factor of all crystals and quasicrystals contains ``Bragg peaks,'' i.e., Dirac delta functions \cite{chaikin2000principles}. Of all possible periodic and quasiperiodic configurations, the structures producing the largest-radius zone around $\mathbf k=\mathbf 0$ devoid of Bragg peaks are the triangular and body-centered cubic lattices in 2D and 3D, respectively. However, if we set $\chi>0.9068\ldots$ in 2D or $\chi>0.9873\ldots$ in 3D, then even these two structures have some Bragg peaks that fall inside the $|\mathbf k|<K$ range; and so do all other crystal and quasicrystal structures \cite{torquato2015ensemble}. Thus, any crystalline or quasicrystalline configuration will have potential energy diverging to plus infinity and thus cannot be an energy minimum \footnote{This also includes two dimensional crystalline and quasicrystalline states at positive temperature that do not have perfect Dirac-delta-function-like Bragg peaks. Rather, the peaks have an intrinsic broadening characteristic, but the broadening is rather limited in magnitude so our interaction still has the effect of banishing phonon-displaced crystalline and quasicrystalline structures in two dimensions \cite{mermin1968crystalline}.}.

Third, this model allows us to realize positive shear and bulk moduli. If a structure corresponding to a local minimum in $\Phi(\mathbf r^N)$ (called an ``inherent structure'' in the rest of the paper) is sheared or compressed at zero temperature, then the set of $\mathbf k$ vectors that are consistent with the simulation box changes. A change in these wave vectors then causes a change in $\mathscr S_0$ (since $\mathscr S_0$ is a function of $\mathbf k$), which in turn will change the potential energy. This change is likely positive because the original configuration is an inherent structure. Indeed, in our simulations, we find this perturbation always increases the potential energy and thus the system will resist that perturbation. Therefore, shear and bulk strains cause stresses. However, for an under-constrained system ($\chi<1$), the unconstrained degrees of freedom allow the system to gradually dissipate the stress over time. Therefore, we always employ $\chi>1$ to ensure that stresses are sustained.

The qualitative nature of our combination of two-, three-, and four-body potentials has the effect of assigning an impossibly high potential energy to structures that have long-range periodic or quasiperiodic order. Thus when the resulting arrangements of particles are disordered by virtue of the nature of the targeted structure factor, these two-, three-, and four-body  contributions to the total energy effectively cancel one another at large distances. Specifically, we show in the SI that for perfect glasses, the three- and four-body contributions to the potential energy almost cancel one another in such a  way  as  to  produce  no  infinite-system thermodynamic anomalies: the total energy per particle is an intensive quantity (as quantitatively detailed in the SI) and approaches its infinite-system-size very rapidly as $N$ increases.

\section*{Results}
\subsection*{Perfect-Glass Inherent Structures}

\label{InherentStructures}

\begin{figure}
\begin{center}
\includegraphics[width=0.7\textwidth]{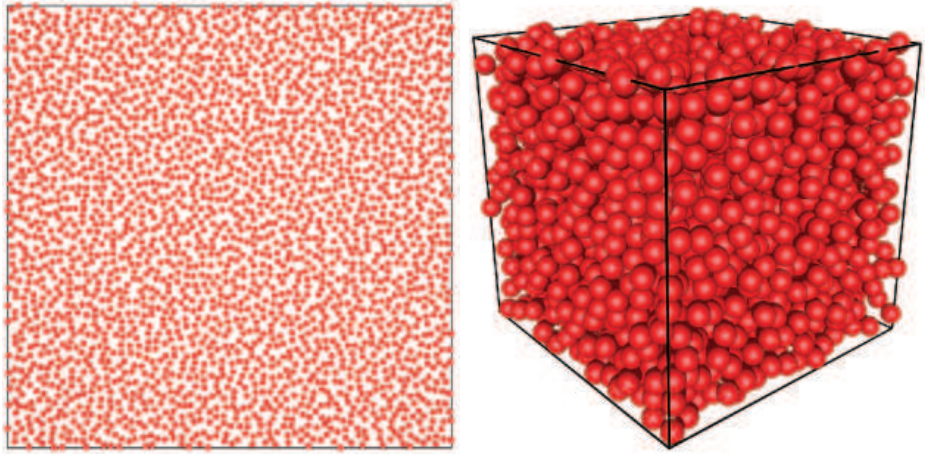}
\end{center}
\caption{Snapshots of perfect glasses with $N=2500$ with perfect-glass potential with parameters $\chi=5.10$, $\alpha=2$, and $\gamma=3$ in 2D (left) and 3D (right). Both of them are clearly disordered.}
\label{Configuration}
\end{figure}

We now quantitatively characterize the structure, elastic moduli, and degree of order of the perfect-glass inherent structures obtained by minimizing the total potential energy, Eq.~(\ref{Potential_Fourier}), starting from random initial configurations of $N=2500$ particles for different parameters $\chi$, $\alpha$, and $\gamma$ in two and three dimensions. Perfect glasses obtained in this way can be regarded as glasses produced by an infinitely rapid quench from infinite temperature to zero temperature because the random initial configuration is equivalent to the infinite-temperature state, and an energy minimization process may be thought of as evolving the system to a state of zero temperature. Examples of perfect glasses in 2D and 3D are shown in Fig.~\ref{Configuration}.

\subsubsection*{Pair Statistics}
\begin{figure}
\begin{center}
\includegraphics[width=0.7\textwidth]{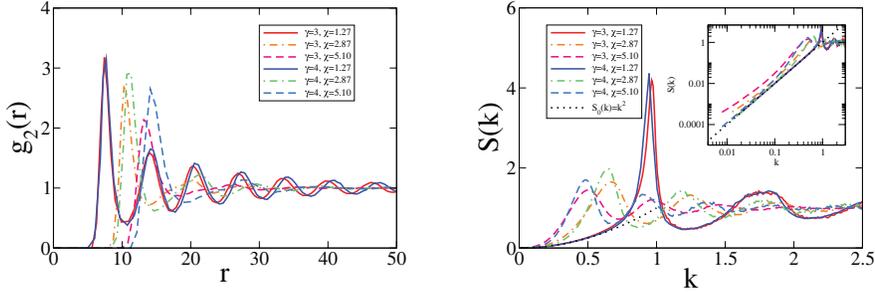}
\end{center}
\caption{Pair correlation functions (left) and structure factors (right) of the perfect glasses in 2D for $\alpha=2$.}
\label{PairStatistics}
\end{figure}

The standard pair correlation function  \cite{chandler1987introduction}, $g_2(r)$, and the angular averaged and ensemble-averaged structure factor, $S(k)$, are together effective descriptors for distinguishing crystals, quasicrystals, disordered hyperuniform systems, and nonhyperuniform systems from one another.
We will restrict ourselves to $\alpha \ge 1$ because, as we will see, this places a lower bound on the rigidity of a perfect glass and is consistent with the MRJ nature of this ideal amorphous state of matter.
These two pair statistics for $\alpha=2$ and different $\chi$'s and $\gamma$'s are shown in Fig.~\ref{PairStatistics}. All $g_2(r)$'s and $S(k)$'s are clearly finite and approaches 1 in the $r \to \infty$ or $k \to \infty$ limit, showing that these structures are neither crystalline nor quasicrystalline. Additionally, $S(k)$ follows the target $\mathscr S_0(k)$ and approaches 0 as $k \to 0$, demonstrating that these structures are hyperuniform. In the SI, we present $S(k)$ for other $\alpha$ and $\gamma$ values and show that $S(k)$ has the same scaling as $\mathscr S_0(k)$ near $k=0$ only if $\gamma>\alpha$. Otherwise, $S(k)$ will deviate from $\mathscr S_0(k)$ and may even appear to saturate at a positive value in the $k \to 0$ limit. 
If so, the resulting system would not be hyperuniform and conform to our definition of a perfect glass.

\subsubsection*{Bulk and Shear Moduli}

\begin{figure}
\begin{center}
\includegraphics[width=0.7\textwidth]{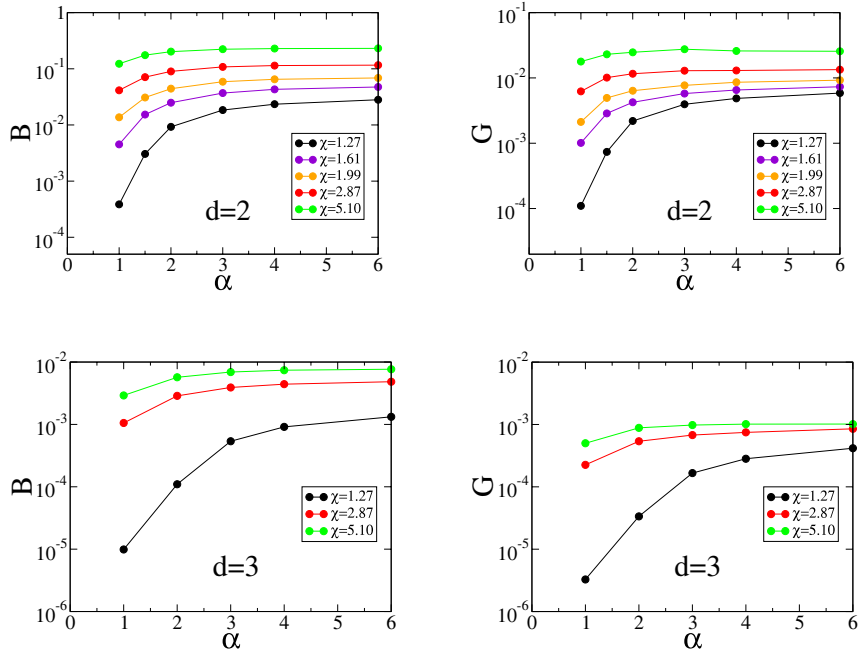}
\end{center}
\caption{Bulk modulus $B$ (left) and shear modulus $G$ (right) versus the exponent $\alpha$  for the inherent structures in 2D (top) and 3D (bottom) for $\gamma=3$ and selected values of $\chi$.}
\label{ElasticConstants}
\end{figure}

Here we show the capacity of a perfect glass to resist both compressive and shear deformations.
The elastic moduli of the inherent structures for $\gamma=3$ are presented in Fig.~\ref{ElasticConstants}. Both moduli increase as $\chi$ or $\alpha$ increases. In all cases, both moduli are positive, clearly showing that our model meets this criterion for a perfect glass. We only present data for $\gamma=3$ for simplicity. It is useful to note that we have also calculated these moduli for $\gamma=2$ or 4 and found the same trend.

Our hyperuniform targeted functional form  $\mathscr S_0(\mathbf k)=|\mathbf k|^\alpha$ generally produces substantially higher elastic constants than those for non-hyperuniform forms; see SI for details.
This correlation between hyperuniformity and improved mechanical rigidity appropriately reflects the MRJ-like nature of the perfect glass and hence stresses the importance of the hyperuniformity criterion.

\subsubsection*{Characterizing the Degree of Disorder}

\begin{figure}
\begin{center}
\includegraphics[width=0.7\textwidth]{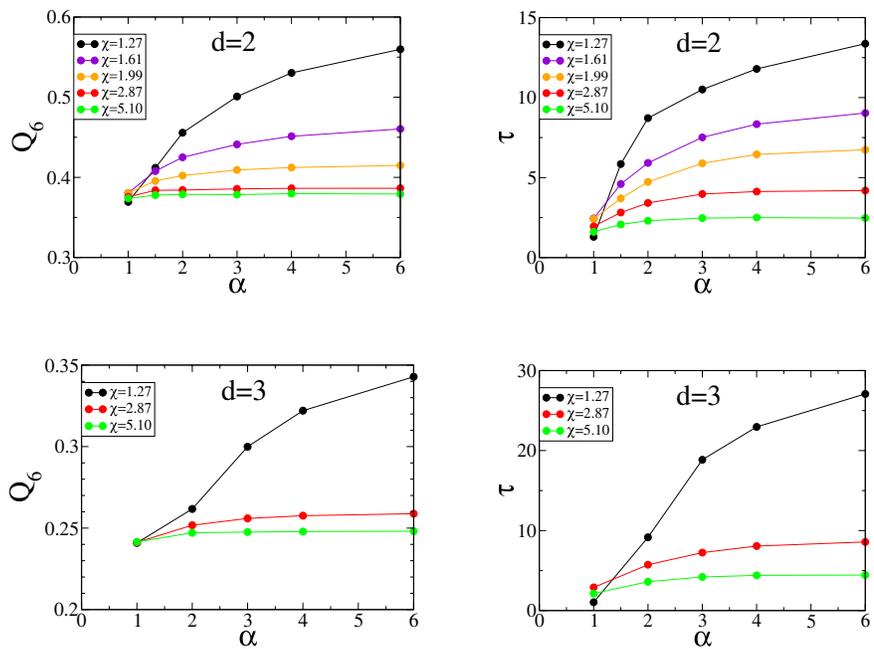}
\end{center}
\caption{Order metrics $Q_{6,local}$ (left) and $\tau$ (right) versus the exponent $\alpha$ for the inherent structures in 2D (top) and 3D (bottom) for $\gamma=3$ and selected values of $\chi$.}
\label{OrderMetric}
\end{figure}

As noted earlier, since perfect glasses are molecular-glass analogs of MRJ sphere packings that are maximally random, we determine here the triplet of parameters ($\chi,\alpha,\gamma$) that produce the most disordered inherent structures according to two order metrics: the ``local'' bond-orientational
parameter $Q_{6,local}$ \cite{kansal2000nonequilibrium, kansal2002diversity} (also denoted $\Psi_6$ in some literature) and the translational order metric $\tau$ \cite{torquato2015ensemble}, which are defined in  the Methods section.
We present the order metrics $Q_{6,local}$ and $\tau$ of the inherent structures in Fig.~\ref{OrderMetric}. Here we want to determine at what value of $\alpha$ is a perfect glass most disordered according to these order metrics. Again, we present data only for $\gamma=3$ for simplicity, but have found that results for $\gamma=2$ or 4 behave similarly. The local bond-orientational order $Q_{6,local}$ measures the degree to which the local environments of particles resemble regular hexagons (in 2D) or regular icosahedra (in 3D); it can vary from 0 (disordered) to 1 (perfect hexagonal order) in 2D or from 0 (disordered) to $0.663...$ (perfect icosahedral order) in 3D. Our relatively low $Q_{6,local}$ values in 3D indicate that our interaction does not favor icosahedral local configuration. This demonstrates that our approach of frustrating crystallization is fundamentally different from the previous approach of encouraging icosahedral order \cite{dzugutov1992glass}. 

\begin{figure}
\begin{center}
\includegraphics[width=0.45\textwidth]{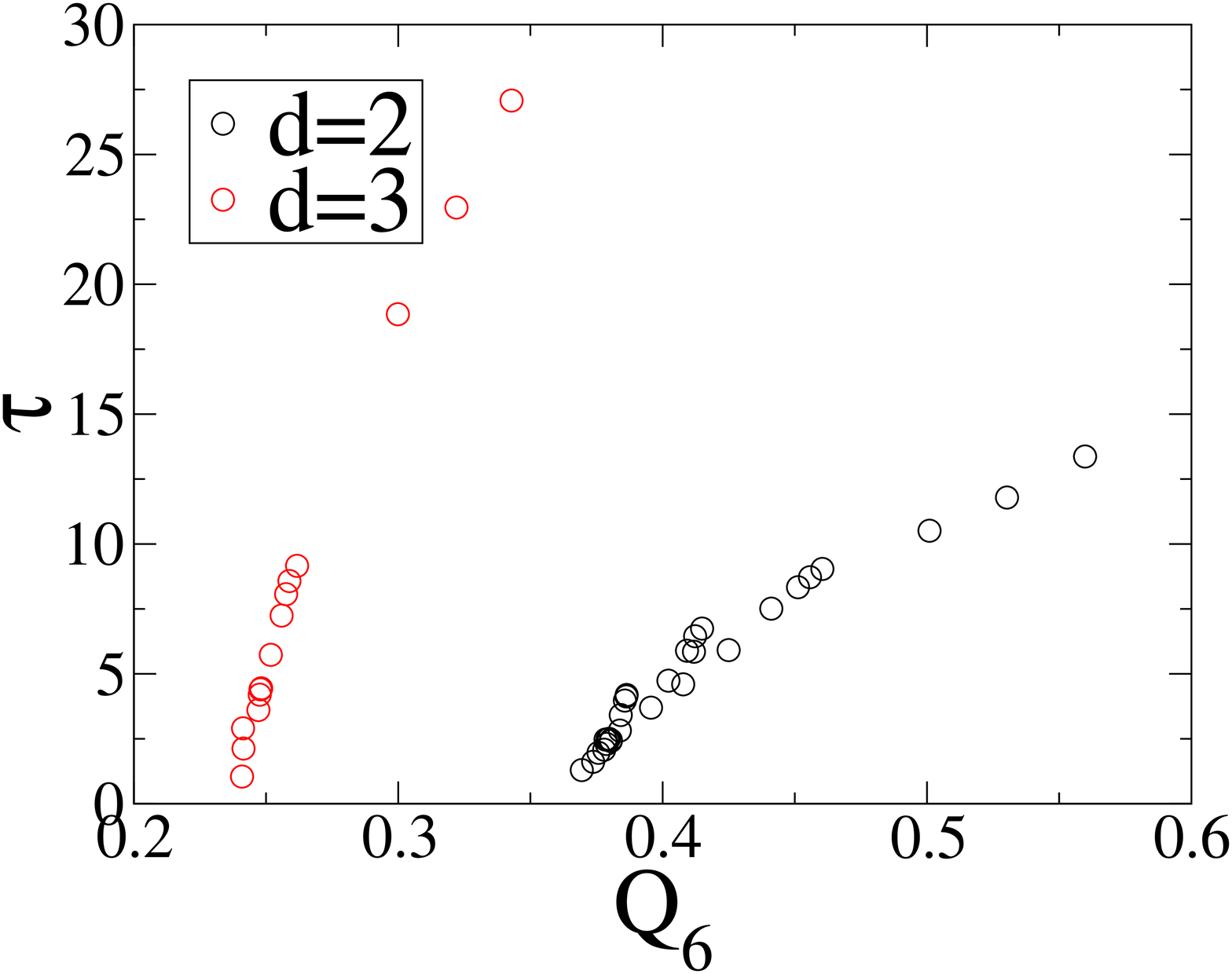}
\end{center}
\caption{Translational order metric $\tau$ versus $Q_{6,local}$ for all previously used $\chi$'s and $\alpha$'s in 2D and 3D for $\gamma=3$.}
\label{OrderMetric2}
\end{figure}

As detailed in the Methods section, $Q_{6,local}$ measures only local orientational order, while the translational order metric, $\tau$, takes into account both short-range order and long-range order. Nevertheless, $\tau$
shows the same trend as $Q_{6,local}$: Perfect glasses with the lowest $\alpha$ and highest $\chi$ have the lowest $\tau$. In fact, we plot $\tau$ versus $Q_{6,local}$ for different $\chi$'s and $\alpha$'s in 2D and 3D for $\gamma=3$ in Fig.~\ref{OrderMetric2} and find that these two order metrics are strongly correlated. Our results for $Q_6$ and $\tau$ are consistent with the qualitative conclusions of Ref.~\onlinecite{uche2006collective}, which reported that increasing $\alpha$ resulted in configurations that increasingly appeared to be more ordered.

While we have shown that the perfect glass is indeed a molecular analog of MRJ, the former is considerably richer. Whereas a perfect glass can have a wide range of degrees of order and elastic moduli, MRJ states, by construction, can only be maximally random subject to the strict jamming condition, which endows them with infinite elastic moduli \cite{torquato2003breakdown}.  
We have shown that as $\alpha$ decreases, the elastic moduli decrease. For this reason and to maintain the analogy with MRJ, we restrict the minimum of $\alpha$ to be 1 so as to bound the elastic moduli from below.
We also must restrict $\alpha<\gamma$ because $S(k)$ demonstrates indisputable hyperuniformity only in such cases.

In summary, we have shown that under the two constraints that $\alpha \ge 1$ and $\alpha<\gamma$, the inherent structures of our potential are clearly disordered, hyperuniform, possess positive shear and bulk moduli, and therefore conform to our definition of a perfect glass. We also note that the lowest value of $\alpha$, equal to unity, produces the lowest order and lowest elastic moduli among all of the cases that we have studied. This behavior is consistent with the fact that MRJ, which are maximally disordered subject to jamming constraint, also have a small-wavenumber scaling of $S(k)\sim k^1$. It has been established that decreasing the exponent $\alpha$ in the small-wavenumber scaling $S(k)\sim k^\alpha$ in many-particle systems is associated with greater disorder and that sublinear scaling ($\alpha<1$) induces clustering among the particles \cite{zachary2011anomalous} and is therefore inconsistent with strict jamming in the case of hard spheres \cite{zachary2011hyperuniformity}.

\subsection*{Simulated Annealing}
\label{SimulatedAnnealing}

The preceding section focused on the inherent structures obtained from random initial configurations, which correspond to glasses produced by an infinitely rapid quench. However, perfect glasses meeting our definition should remain disordered even after annealing with a finite, slow cooling rate. Here we study the behavior of our system under slow annealing by performing canonical ensemble (constant temperature and volume) molecular dynamics (MD) simulations \cite{frenkel1996understanding} and gradually decreasing the temperature.
We have performed such an annealing for a 2D system of $N=400$ particles with parameters $\chi=5.1$, $\alpha=1$, and $\gamma=3$. During the annealing process, the potential energy remains continuous with respect to temperature, suggesting that there is no first-order phase transition. 
As we will mention in the Methods section, the configuration starts to vibrate around a single inherent structure when $k_B T$ drops below $0.3$, suggesting that the glass transition temperature, $T_g$, for this system at this cooling rate is around $0.3/k_B$.
The final configuration is disordered, verifying that our system does not crystallize even under slow cooling.

It is worth noting that after slow annealing and a subsequent energy minimization, the final configuration has potential energy per particle $\Phi/N=2.920$, which is not much lower than that of the previously obtained inherent structures, for which, $\Phi/N=2.971\pm0.014$ at the same system size under the same interaction. This may suggest that most of the local energy minima of the potential energy surface are not much higher than the ground state energy. 

\section*{Conclusions and Discussion}

We have found a family of interactions that can produce perfect glasses, i.e., hyperuniform glasses with positive bulk and shear moduli, in systems that possess no crystalline or quasicrystalline energy minima. We have demonstrated that the inherent structures (structures obtained by infinitely rapid cooling from infinite temperature to zero temperature) of these interactions are disordered, hyperuniform, and have positive bulk and shear moduli for parameters $1\le \alpha < \gamma$ and $\chi>1$. The lowest $\alpha$, equal to unity, results in the lowest degree of order, although a priori there was no reason to expect that maximum disorder would arise when $\alpha$ is minimized. 
 We have also performed a slow simulated annealing on a perfect-glass system and found no first-order phase transition.

Our interactions are designed in Fourier space and completely eliminate crystal and quasicrystal formation. As detailed in Ref.~\onlinecite{torquato2015ensemble}, for $\chi>0.9068\ldots$ in 2D or $\chi>0.9873\ldots$ in 3D, any crystal or quasicrystal must produce Bragg peaks in the constrained ($|\mathbf k|<K$) range. Such Bragg peaks would make the potential energy infinite. Therefore, crystals and quasicrystals cannot be energy minima  \footnote{{ It is intersting to note that our approach to ensure glass formation (eliminating crystals and quasicrystals) is in sharp contrast with the reason why diboron trioxide tends to vitrify. The latter tends to vitrify because of a high energy degeneracy of multiple crystalline structures \cite{ferlat2012hidden}.}}.
Since our perfect glasses are not metastable with respect to a crystal structure, there is no Kauzmann entropy crisis that led to the conjectured existence of ``ideal glasses'' in the conventional Kauzmann picture \cite{debenedetti2001supercooled, stillinger2013glass}. The latter is defined completely differently from the perfect glasses in this work. 

All available understanding indicates that with only isotropic two-body interactions, crystalline ground states inevitably occur. However, by adding suitable three- and four-body interactions to appropriate two-body interactions, we show for the first time that crystals and quasicrystals can be completely prevented for any range of temperatures down to absolute zero and thus ensures by construction thermodynamically stable glassy states.
The collective-coordinate procedure that we are using to target perfect glass behavior cannot be simplified to the extent of reducing the interaction character below at least four-body interactions.
Because the procedure is general, this suggests that a perfect glass cannot be created with two- and three- body interactions depending only on scalar distances alone. This could explain why previous attempts to produce such an ideal state of matter have not been successful. Another observation that suggests the necessity of four-body interactions is that the analytical form of our four-body interaction, Eq.~(\ref{pot_v4}), appears to strongly penalize long-range bond orientational order, and thus prevent crystallization.
Without the four-body interaction, one might be able to design a pair interaction that reproduces the pair correlation function of perfect glasses at a particular temperature (for example, by imitating Ref.~\onlinecite{gerold1986determination}). However, crystallization cannot be prevented if the temperature is lowered down to absolute zero.

It is instructive to compare and contrast the perfect-glass potential with the well-known classical rigidity theory of Phillips and Thorpe \cite{thorpe1983continuous, micoulaut2016concepts}. This theory applies to glasses with covalent interactions, and states that covalent bonds between atoms in a glass impose constraints on the atomic positions and that the conditions for glass formation will be optimal if the number of constraints is equal to the number of degrees of freedom of the atoms.
By contrast, the perfect-glass picture involves overconstraining the system (i.e., having more constraints than the degrees of freedom), which occurs when $\chi>1$, to ensure positive elastic moduli. Moreover, the Phillips-Thorpe theory states that normal glasses tend to crystallize if the number of constraints is much larger than the number of degrees of freedom, but a perfect glass will never crystallize even for large $\chi$ values. Besides these differences, perfect glasses have other important distinctive features, e.g., hyperuniformity and complete prevention of crystallization. Lastly, the isotropic perfect-glass interaction is also very different from the directional covalent-bond interactions that the Phillips-Thorpe theory assumes, and a perfect glass is achievable with identical particles.

It is worth noting that our model systems can maintain hyperuniformity even at positive temperatures. Eqs.~(\ref{Potential_Fourier})-(\ref{weight}) suggests that, with sufficiently high $\gamma$, any nonhyperuniform structure will have infinite energy, and therefore have zero probability of appearing at a finite positive temperature. 
We analyzed the intermediate configurations from the annealing simulation at $T=10$ and indeed found hyperuniformity.
This feature contrasts with other interactions that have perfect crystalline, and therefore hyperuniform, ground states but lose hyperuniformity at any positive temperature due to phonon excitations (e.g., Lennard-Jones interaction).  

The well-known compressibility relation from statistical mechanics \cite{hansen1990theory} usually provides 
some insights about the relationship between temperature $T$ and hyperuniformity
for equilibrium systems in the infinite-system-size limit at number density $\rho=N/V$:
\begin{equation}
\lim_{k \to 0} S(k)=\rho k_B T \kappa_T.
\label{CompressibilityRelation}
\end{equation}
We see that 
in order to have a hyperuniform equilibrium
system at positive $T$ that obeys this relation, the isothermal compressibility, $\kappa_T=1/B$, must be zero; i.e.,
the system must be incompressible \cite{zachary2011anomalous, torquato2015ensemble} (see Refs. \onlinecite{torquato2003local} and \onlinecite{torquato2008point}
for some examples). 
As stated in the previous paragraph, equilibrium systems of particles interacting with the perfect-glass potential at positive temperature (e.g., liquids) are hyperuniform. Does this mean they are also incompressible ($B = \infty$)? Our initial study of perfect glasses in isothermal-isobaric ensembles suggests that they are not incompressible. Thus, the compressibility relation is violated. The reason for this violation is that there are actually two subtly different compressibilities: the ``internal'' one and the ``external'' one. If one divides a large system into two halves, compresses one half and decompresses the other half while keeping the total volume constant, the restoring force is related to the internal compressibility. However, if one compresses or decompresses the entire system, causing a volume change, the change in pressure is related to the external compressibility. Normally, the internal compressibility is equal to the external one and thus the compressibility relation holds. However, for perfect glass systems, since the potential energy explicitly penalizes long-wavelength internal density fluctuations but not external volume change, the internal compressibility is zero while the external one is still positive and the compressibility relation no longer holds. A novel consequence of having zero internal compressibility is that the non-relativistic speed of sound is infinite.

Concerning the violation of the compressibility relation, it is interesting to note that we previously have studied Dzugutov glasses and Lennard-Jones glasses, which violate the same relation because they are not in equilibrium, and introduced the following ``non-equilibrium index'' \cite{marcotte2013nonequilibrium}:
\begin{equation}
X=\frac{\lim_{k \to 0} S(k)}{\rho k_B T \kappa_T}-1.
\end{equation}
If a system is not in equilibrium and thus violates the compressibility relation, $X$ would be non-zero. However, systems of particles interacting with the perfect-glass potential, even in equilibrium, would still have a non-zero $X$. 

Our perfect-glass model has the unique feature of not being metastable with respect to any crystalline or quasicrystalline states. We believe these features can open up a variety of possibilities. Without the worry of crystallination and with the help of faster computers in the future, one would be able to perform extremely long simulations to study glass dynamics. It would be an interesting future project to
study the kinetics of glass formation as a function of temperature and density.
It would also be interesting to see whether or not the ground states of the perfect-glass interaction have vanishing configurational entropy per particle. If so, this would be the first example of this conjectured ``ideal glass'' \cite{stillinger2013glass, debenedetti2001supercooled}. As an initial attempt to see whether the perfect glass has this attribute, we have reduced the number of particles to 10, a computationally manageable number, and performed simulated annealing 150 times and obtained lowest-energy states four times. Presumably, these are the ground states. The ground
state configurations obtained this way cannot be related to each other via
simple translation, rotation, or inversion, and hence are
geometrically degenerate. However, this is not a definitive evidence
that perfect-glass systems are not ideal glasses at
absolute zero. It is still possible that, although the ground state is degenerate,
the number of degenerate structures does not scale exponentially
with the system size. If so, perfect-glass systems could still have vanishing
configurational entropy per particle at absolute zero in the infinite-system-size limit. Faster computers in the future should allow the same investigation for larger systems in order to ascertain the scaling of the number of degenerate ground states with the system size for perfect-glass systems.

Concerning the first criterion of perfect glasses (hyperuniformity), we note in passing that real polymers \cite{hess2006long} as well as polymer models \cite{eilhard1999spatial, xu2016entropy} have succeeded in approaching hyperuniformity.
It remains to be seen whether the remaining two criteria can be approached by novel polymer systems or suitably defined theoretical models of polymers.
It is also worth noting that polymer systems are known to involve high-order interactions beyond two-body
terms \cite{bolhuis2001many, dijkstra2002entropic, watzlawek1999phase}, which, as we discussed earlier, are likely required to create perfect glasses.

There is a broader class of mathematical models as those for which $\lim_{\mathbf k \to \mathbf 0} \mathscr S_0(\mathbf k) \neq 0$ or $\lim_{\mathbf k \to \mathbf 0} {\tilde v}(\mathbf k) \neq +\infty$. Generally, they would produce nonhyperuniform glasses and if so, would not conform to our definition of perfect glasses. 
Nevertheless, such models still completely eliminate crystalline and quasicrystalline energy minima and therefore merit future mathematical analyses and numerical studies. This is in contrast to a study in which a similar type of potential was added to a Lennard-Jones interaction in order to inhibit crystallization  \cite{di2000off, angelani2002quasisaddles}, but the functional form employed prevented that goal from being accomplished \cite{angelani2002quasisaddles}. 

\section*{Methods}

\label{SimulationDetails}

We generate inherent structures of perfect-glass potentials by the following procedure: Starting from initial configurations of $N=2500$ particles in which each particle's position is generated randomly and independently, we minimize the potential energy, Eq.~(\ref{Potential_Fourier}), first using the low-storage BFGS algorithm \cite{nocedal1980updating, liu1989limited, nlopt} and then using the MINOP algorithm \cite{dennis1979two}. Such a combination of the two minimization algorithms maximizes both efficiency and precision \cite{zhang2015ground}. After energy minimization, the norm of the gradient of potential energy is less than $10^{-13}$ (in dimensionless units, similarly hereinafter). The simulation box shape is square in 2D and cubic in 3D. Since $K$ and $N$ are fixed, we adjust $\chi$ by changing the simulation box size. We choose side lengths $L=$400, 450, 500, 600, and 800 for $\chi=$1.27, 1.61, 1.99, 2.87, and 5.10, respectively, in 2D and $L=$104.1, 136.6, and 165.4 for $\chi=$1.27, 2.87, and 5.10, respectively, in 3D. For a 2D case in which $\chi=5.10$, $\alpha=2$, and $\gamma=3$, we also generated inherent structures in a rhombic simulation box with a $60^\circ$ interior angle and have found no statistically significant difference in the resulting pair correlation function, structure factor, and elastic constants, verifying that our results are not sensitive to the shape of the simulation box. For each combination of $\chi$, $\alpha$, and $\gamma$ in both 2D and 3D, we generated between 10 and 100 inherent structures, depending on the energy minimization speed of the specific case.

To demonstrate that our perfect glasses have positive bulk moduli ($B$) and shear moduli ($G$), we have also calculated these elastic moduli of the inherent structures by incurring a small ($10^{-6}$) strain, minimizing the potential energy within the deformed simulation box, and then calculating the stress. The calculated elastic constants are then averaged over different directions of strains and stresses and different configurations.

Since perfect glasses are molecular-glass analogs of hard-sphere MRJ packings that are maximally random, we are interested in finding the triplet of parameters  ($\chi,\alpha,\gamma$) that produce the most disordered inherent structures according to certain order metrics. We have calculated two order metrics: $Q_{6,local}$ and $\tau$.
In 2D, $Q_6$ is 
defined for a given particle $q$, as
\begin{equation}
Q_6=\left|\frac{1}{N_p} \sum_p \exp(6i\theta_{\mathbf r_{pq}})\right|,
\end{equation}
where the summation is over all neighbor particles whose Voronoi cells share an edge with particle $q$'s cell, 
$N_p$ is the number of such neighbors, and $\theta_{\mathbf r_{pq}}$ is the angle between $\mathbf r_{pq}=
\mathbf r_q-\mathbf r_p$ and a reference direction. In 3D, $Q_6$ is defined as
\begin{equation}
Q_6=\sqrt{\frac{4\pi}{13}\sum_{m=-6}^6 \left|\frac{1}{N_p}\sum_p Y_{6m}(\theta, \phi)\right|^2},
\end{equation}
where the summation is over all neighbor particles whose Voronoi cells share a face with particle $q$'s cell, 
$N_p$ is the number of such neighbors, $Y_{lm}$ is the spherical harmonic function, and $\theta$ and $\phi$ 
represent colatitude and longitude of $\mathbf r_{pq}$. These bond-orientational parameters are the local 
versions of the ones introduced in Ref.~\onlinecite{steinhardt1983bond}. In both dimensions, $Q_{6,local}$ is 
an average of $Q_6$ over all particles in all configurations.

While $Q_{6,local}$ only measures local orientational order, the following 
translational order metric \cite{torquato2015ensemble}:
\begin{equation}
\tau=\rho \int_0^\infty [g_2(r)-1]^2 d\mathbf r = \frac{1}{(2\pi)^d \rho}\int_0^\infty [S(k)-1]^2 d\mathbf k,
\label{tau}
\end{equation}
takes into account both short-range order and long-range order by measuring the degree to which the pair statistics [$g_2(r)$ and $S(k)$] deviate from those of an ideal gas on all length scales.
As Eq.~(\ref{tau}) shows, $\tau$ can be computed from either $g_2(r)$ or $S(k)$. Parseval's theorem guarantees that these two approaches yield the same value of $\tau$ in the infinite-system-size limit. However, they can give slightly different results for our finite-sized systems, and hence provide
a self-consistency check
on its evaluation in a simulation.
Although Eq.~(\ref{tau}) involves infinite integrations, they can be truncated since both $g_2(r)$ and $S(k)$ decay and approach 1 rapidly in the $r\to\infty$ or $k\to\infty$ limit.
In our calculation, the integration is truncated at $r_{cut}=200$ in 2D and $r_{cut}=50$ in 3D or $k_{cut}=6$ in both dimensions.
As we will show in the SI, $\tau$ calculated from both approaches agree well, verifying that our $g_2(r)$ and $S(k)$ are consistent and our integration truncation is appropriate.

To demonstrate that perfect glasses cannot crystallize, we have also performed molecular-dynamics-based simulated annealing of the perfect glass potential using the velocity Verlet algorithm \cite{frenkel1996understanding}. The temperature is controlled by resetting a randomly chosen particle's velocity to a random velocity, drawn from Boltzmann distribution, every 10 time steps. The scaled temperature, $k_BT$, starts at 10 in dimensionless units and decreases as prescribed by Eq.~(6) of Ref.~\onlinecite{nourani1998comparison}. In evaluating that equation, we use the relaxation time of the potential energy $\Phi$ as an estimate of the relaxation time of the system and use the scaling parameter in that Eq.~(6) $v_s=0.6$.
The integration time step $\Delta t$ is adjusted continuously so that the change in total energy every 50 time steps is between 0.0025\% and 0.01\% when velocity resetting is switched off. In our simulation $\Delta t$ changed from 0.05 at $k_BT=10$ to 0.18 at $k_BT=0.23$. As $k_BT$ dropped below about 0.3, the configuration started vibrating around a single inherent structure and we thus ended the simulation. The time length of the entire MD simulation is $t=3.04\e6$ in dimensionless unit.

\section*{Appendix: Perfect-glass potential in the direct space}
\label{decomposition}
As we mentioned in the ``Perfect Glass Potentials'' section, the perfect-glass potential, Eq. ~(\ref{Potential_Fourier}), can be decomposed into a sum of two-, three-, and four-body contributions in the direct space. We present explicit formulas for these contributions here. 
We provide visulizations of these individual two-, three-, and four-body contributions in the SI.

The total potential energy for $N$ particles in a fundamental cell under periodic boundary condition is given by \cite{uche2006collective}
\begin{equation}
\Phi(\mathbf r^N) = \sum_{0<|\mathbf k|<K} {\tilde v}(\mathbf k) [\mathscr  S(\mathbf k)-\mathscr S_0(\mathbf k)]^2  = \sum_{l < m < n < p} v_4(\mathbf r_l, \mathbf r_m, \mathbf r_n, \mathbf r_p) + \sum_{l < m < n } v_3(\mathbf r_l, \mathbf r_m, \mathbf r_n)  + \sum_{l < m } v_2(\mathbf r_l, \mathbf r_m) + v_0,
\end{equation}
where
\begin{equation}
v_4(\mathbf r_l, \mathbf r_m, \mathbf r_n, \mathbf r_p) = \frac{8}{N^2} \sum_{0<|\mathbf k|<K} {\tilde v}(\mathbf k) [
\cos(\mathbf k \cdot \mathbf r_{lm}) \cos(\mathbf k \cdot \mathbf r_{np}) 
+\cos(\mathbf k \cdot \mathbf r_{ln}) \cos(\mathbf k \cdot \mathbf r_{mp}) 
+\cos(\mathbf k \cdot \mathbf r_{lp}) \cos(\mathbf k \cdot \mathbf r_{mn}) 
], 
\label{pot_v4}
\end{equation}
\begin{equation}
v_3(\mathbf r_l, \mathbf r_m, \mathbf r_n) = \frac{8}{N^2} \sum_{0<|\mathbf k|<K} {\tilde v}(\mathbf k) [
\cos(\mathbf k \cdot \mathbf r_{lm}) \cos(\mathbf k \cdot \mathbf r_{ln}) 
+\cos(\mathbf k \cdot \mathbf r_{lm}) \cos(\mathbf k \cdot \mathbf r_{mn}) 
+\cos(\mathbf k \cdot \mathbf r_{ln}) \cos(\mathbf k \cdot \mathbf r_{mn}) 
], 
\end{equation}
\begin{equation}
v_2(\mathbf r_l, \mathbf r_m) = \frac{4}{N} \sum_{0<|\mathbf k|<K} {\tilde v}(\mathbf k) \cos(\mathbf k \cdot \mathbf r_{lm}) [1-\mathscr S_0(\mathbf k)+\cos(\mathbf k \cdot \mathbf r_{lm})/N],
\end{equation}
and
\begin{equation}
v_0=\sum_{0<|\mathbf k|<K} {\tilde v}(\mathbf k) [1-\mathscr S_0(\mathbf k)]^2.
\end{equation}

\section*{Acknowledgements}

We thank Steven Atkinson for his careful reading of the manuscript. This work was supported by the
National Science Foundation under Grant No. DMS-1211087.

\section*{Author contributions statement}

S.T. conceived the research,  devised the methods, G. Z. performed the simulations, 
G. Z., F. S. and S. T. performed analysis, and  G. Z., F. S. and S. T. wrote the paper.

\section*{Additional information}

The authors declare no competing financial interests.

\end{document}